\documentclass[a4paper,11pt]{article}
\usepackage{pos}
\usepackage{annotate-equations}
\usepackage{accents}

\renewcommand{\d}{\mathrm{d}}
\newcommand{\tr}[1]{\mathrm{Tr}\left\{#1\right\}}

\definecolor{customblue}{HTML}{3c9bb3}
\definecolor{customred}{HTML}{9d1700}
\definecolor{customyellow}{HTML}{e0ad04}

\definecolor{customgreen}{HTML}{117877}
\definecolor{custompink}{HTML}{c35861}

\definecolor{starrymain}{HTML}{3B5B65}
\definecolor{starrysecond}{HTML}{719593}

\title{Heavy quark $\kappa$ and jet $\hat{q}$ transport coefficients in the Glasma early stage of heavy-ion collisions}
\ShortTitle{Heavy quark $\kappa$ and jet $\hat{q}$ transport coefficients in the Glasma}

\author*[a,b]{Dana Avramescu}
\author[c]{Virgil Băran}
\author[d, e]{Vincenzo Greco}
\author[f]{Andreas Ipp}
\author[f]{David Müller}
\author[d, e]{Marco Ruggieri}

\affiliation[a]{Department of Physics, University of Jyväskylä\\
P.O. Box 35, 40014 University of Jyväskylä, Finland}

\affiliation[b]{Helsinki Institute of Physics\\
P.O. Box 64, 00014 University of Helsinki, Finland}

\affiliation[c]{Faculty of Physics, University of Bucharest\\
Atomiștilor 405, Măgurele, Romania}

\affiliation[d]{Department of Physics and Astronomy, University of Catania\\
Via S. Sofia 64, I-95123 Catania, Italy}

\affiliation[e]{INFN-Laboratori Nazionali del Sud\\
Via S. Sofia 62, I-95123 Catania, Italy}

\affiliation[f]{Institute for Theoretical Physics, TU Wien\\
Wiedner Hauptstraße 8, A-1040 Vienna, Austria}

\emailAdd{dana.d.avramescu@jyu.fi, virgil.baran@unibuc.ro, greco@lns.infn.it, ipp@hep.itp.tuwien.ac.at, dmueller@hep.itp.tuwien.ac.at, marco.ruggieri@dfa.unict.it}

\abstract{We study the impact of the Glasma fields, used to describe the very early stage of heavy-ion collisions, on the transport of hard probes, namely heavy quarks and jets. We perform numerical simulations of the strong classical fields using techniques from real-time lattice gauge theory. The resulting fields are used as background for the classical transport of ensembles of particles, described by Wong's equations. For this purpose, we develop a numerical solver for the transport of the probes, based on colored particle-in-cell methods. 

We focus on the dynamics of heavy quarks and jets in the classical colored fields. To quantify the effect of the Glasma, we extract the momentum broadening of hard probes and evaluate the anisotropy transfer from the Glasma to the probes. Lastly, we evaluate the heavy quark $\kappa$ and jet $\hat{q}$ transport coefficients in the Glasma, which turn out to be large and exhibit a peak, irrespective of the particle initialization.}

\FullConference{HardProbes2023\\
 26-31 March 2023 \\
 Aschaffenburg, Germany\\}

\begin{document}
\maketitle

\section*{Introduction}
\textbf{Framework.} A widely accepted approach in our field is to model relativistic heavy-ion collisions as a sequence of distinct stages. Each stage is formulated within a suitable effective theory. In this work, we are interested in the initial stage and use the Color Glass Condensate (CGC) theory \cite{Gelis:2010nm} to describe it. The CGC lies at the high energy limit of Quantum Chromodynamics. It is based on the observation that, as opposed to a nucleus at rest which consists only of valence quarks, at high energy, the wavefunction of a high-energy nucleus quickly becomes dominated by gluons. Large occupation numbers characterize the associated gluon fields. Thus, we may treat them as classical. The classical Yang-Mills fields which arise from the collision of two such energetic nuclei are named the Glasma \cite{Lappi:2006fp}. 

\textbf{Literature.} In the literature, there has been increasing interest in studying whether the hard probes, which are produced early in a collision, are affected by the Glasma initial stage. A pioneer work \cite{Ruggieri:2018rzi} first studied the heavy quark diffusion in Glasma, followed by investigating the effect of the Glasma stage on the $R_{AA}$ and $v_2$ puzzle \cite{Sun:2019fud}. In another approach, the jet transport coefficient $\hat{q}$ was extracted from Wilson loops on the lattice for light-like jets \cite{Ipp:2020mjc, Ipp:2020nfu}. The intriguing result was that $\hat{q}$ in the Glasma is large and exhibits a peak. Lattice results for static heavy quark transport coefficient $\kappa$ in an over-occupied gluon plasma \cite{Boguslavski:2020tqz} complemented this observation, revealing that the momentum broadening increases rapidly in the early stage. Further analytical studies \cite{Carrington:2021dvw, Carrington:2022bnv} supported the picture of large transport coefficients for both heavy quarks and jets in the Glasma. Lastly, recent results using effective kinetic theory, within the bottom-up thermalization scenario, \cite{Boguslavski:2023alu, Boguslavski:2023fdm} seamlessly fill the gap between the large $\kappa$ and $\hat{q}$ in the Glasma and smaller values from the subsequent hydrodynamics stage.

\textbf{This study.} All the aforementioned studies suggest that the Glasma has a significant impact on the early hard probes. In our study \cite{Avramescu:2023qvv}, we manage to surpass certain approximations done in these previous works, at the level of either Glasma or particle transport. Our refinements consist of simulating the Glasma fields on a SU(3) lattice and implementing full particle dynamics. More concisely, we employ classical lattice gauge theory to solve the Glasma fields. These fields are then used as background for the propagation of particles. Their dynamics are evolved from Wong's equations and are numerically solved using the colored particle-in-cell method. In our particle solver, we can consider the effect of finite quark masses, formation times and initial momenta.

\section{Hard probes in Glasma}

\textbf{Glasma.} The Glasma is constructed within the CGC framework. Partons from a nucleus carry different amounts of the total momentum, quantified through the Bjorken $x$ variable. CGC is based on the high-energy separation of scales between soft, {\color{customgreen}small-$x$} partons and hard {\color{custompink}large-$x$} partons. The {\color{custompink}large-$x$} partons, mostly valence quarks, act as {{\color{custompink}color sources} for the many {\color{customgreen}small-$x$} partons, the {\color{customgreen}classical gluon fields}, as described by the Yang-Mills equation

\vspace{1em}
\renewcommand{\eqnhighlightheight}{\vphantom{\mathcal{D}_\mu}\mathstrut}\begin{equation}\label{eq:cym}
\big(\eqnmark[starrymain]{dmu}{\mathcal{D}_\mu}\hspace{-0.2cm}\eqnmark[starrysecond]{fmunu}{F^{\mu\nu}}\big)\big[\hspace{-0.1cm}\eqnmark[customgreen]{amu}{A^\mu}\hspace{-0.1cm}\big]=\eqnmark[custompink]{jnu}{J^\nu}.
\end{equation}
\annotate[yshift=0.7em]{above, left}{dmu}{covariant derivative}
\annotate[yshift=0.7em]{above}{fmunu}{field strength tensor}
\annotate[yshift=-0.5em]{below, left}{amu}{gluon gauge field}
\annotate[yshift=-0.5em]{below, right}{jnu}{nucleus color current}

The color component $a$ of the current of a high-energy nucleus moving along the light-cone direction $x^\pm$ is ${\color{custompink}J}^{{\color{custompink}\mu},a}(x^-, x^+, \undertilde{x}) \propto \delta^{\mu\pm}\delta(x^\mp)\rho^a(\undertilde{x})$, where $\undertilde{x}\equiv(x, y)$. We further employ the MV model which assumes that color charges $\rho^a$ in large nuclei are stochastic variables. Their two-point function $\langle \rho^a \rho^a \rangle \propto Q_s^2$ is governed by $Q_s$ the saturation momentum, namely the scale at which the gluon occupation number starts to saturate. Besides numerical lattice parameters, it is the only physical parameter of the Glasma. 

After the collision of two such thin recoilless nuclei, the Glasma fields are produced in the forward light cone. In our approach, we switch to Milne coordinates ($\tau$, $\eta$), where $\tau\equiv\sqrt{2 x^+x^-}$ is the Milne proper time and $\eta\equiv\ln(x^+/x^-)/2$ the spatial rapidity, and impose boost-invariance of the Glasma fields, namely $A^\mu(x)=A^\mu(\tau, \undertilde{x})$. This approximation enables us to numerically solve the Glasma field equations on the lattice, discretized in terms of gauge link and plaquette variables. 

One important feature of the Glasma fields is that their properties are primarily dictated by the saturation scale $Q_s$. At late times, of the order $\delta\tau_\mathrm{free-stream}\sim 1/Q_s$, the fields enter the free-streaming regime and become more dilute. The color electromagnetic fields arrange themselves in correlation domains of typical size $\delta\undertilde{x}_\mathrm{\,corr}\sim 1/Q_s$. Moreover, because of the beam axis, the field configurations are highly anisotropic. All these features will be imprinted in the subsequent dynamics of the probes in the Glasma fields.

\textbf{Probes in Glasma.} The dynamics of particles in Yang-Mills fields are governed by Wong’s equations

\renewcommand{\eqnhighlightheight}{\vphantom{x}}
\begin{equation}
    \label{eq:wong}
    \frac{\d}{\d\hspace{-0.1cm}\eqnmark{tau}{\boldsymbol{\tau}}\hspace{-0.2cm}}\eqnmark[customblue]{xmu}{x^\mu}=\frac{{\color{customred}p^\mu}}{\eqnmark{m}{m}},\qquad \eqnmark{Ddtau}{\frac{\mathrm{D}}{\d\boldsymbol{\tau}}}\hspace{-0.2cm}\eqnmark[customred]{pmu}{p^\mu}=2\hspace{-0.1cm}\eqnmark{g}{g}\hspace{-0.1cm}\tr{{\color{customyellow}Q}F^{\mu\nu}[\hspace{-0.1cm}\eqnmark[customgreen]{amu}{A^\mu}\hspace{-0.1cm}]}\frac{{\color{customred}p_\nu}}{m},\qquad 
    \underbrace{\frac{\d}{\d\boldsymbol{\tau}}\hspace{-0.1cm}\eqnmark[customyellow]{Q}{Q}\hspace{-0.1cm}=-\mathrm{i}g [{\color{customgreen}A_\mu},{\color{customyellow}Q}]\,\frac{{\color{customred}p^\mu}}{m}}_{\substack{\text{\footnotesize color rotation}\,\rightarrow\,{\color{starrymain}\mathcal{U}}\in\,\mathrm{SU(3)} \\[0.2cm] {\color{customyellow}Q}(\boldsymbol{\tau})=\,{\color{starrymain}\mathcal{U}}(\boldsymbol{\tau},\boldsymbol{\tau}_0){\color{customyellow}Q}(\boldsymbol{\tau}_0)\,{\color{starrymain}\mathcal{U}^\dagger}(\boldsymbol{\tau},\boldsymbol{\tau}_0)}}.
\end{equation}
\annotate[yshift=1.2em]{above}{xmu}{coordinate}
\annotate[yshift=1.2em]{above}{pmu}{momentum}
\annotate[yshift=-0.5em]{below, right}{m}{\tiny mass}
\annotate[yshift=-1.5em]{below, right}{Ddtau}{\tiny covariant derivative}
\annotate[yshift=-1.5em]{below, right}{tau}{\tiny proper time}
\annotate[yshift=-0.7em]{below, right}{g}{\tiny coupling constant}
\annotate[yshift=1.2em]{above}{Q}{color charge}
\annotate[yshift=1.2em]{above, right}{amu}{gauge field}

They describe how the coordinate ${\color{customblue}x^\mu}$, momenta ${\color{customred}p^\mu}$ and color charge ${\color{customyellow}Q}$ of classical particles change while propagating in a background gauge field ${\color{customgreen}A^\mu}$. The momenta evolution contains, in disguise, the color Lorentz force in curvilinear coordinates. The equation involving the charge may be recast in terms of a ${\color{starrymain}\mathcal{U}}\in\mathrm{SU(3)}$ color rotation. In our particle solver, we color rotate the charge with particle Wilson lines constructed from the underlying Glasma lattice gauge links. This procedure is inspired by the colored particle-in-cell method and has the numerical advantage of conserving colored charges algebra invariants. 

The color charges $ Q=Q^a T^a$ are constructed with fixed quadratic $q_2\equiv Q^a Q^a$ and cubic $q_3\equiv d_{abc}Q^a Q^bQ^c$ Casimir invariants, where $T^a$ are SU(3) generators and $d_{abc}$ the symmetric structure constants. The ensemble of classical color components are distributed according to the one-, two- and three-point functions $\langle Q^a \rangle=0$, $\langle Q^a Q^b \rangle =1/2$ and $\langle Q^a Q^b Q^c \rangle=d^{abc}/4$, in the fundamental representation. Our procedure for generating random classical color charges relies on first constructing an initial quark color charge $Q_0=Q_0^a T^a$ with desired $q_2$ and $q_3$ values and then performing random color rotations $Q(\tau_0)=U Q_0 U^\dagger$ with $U \in \mathrm{SU(3)}$ chosen according to the Haar measure. This gives the initial color vector $Q(\tau_0)$ from Eq.~\eqref{eq:wong}. The subsequent color rotations with particle Wilson lines automatically conserve the initially fixed Casimir invariants. More details about classical color algebras and how the color charge equation is numerically solved on the lattice are provided in \cite{Avramescu:2023qvv}.

\section{Key results}

\textbf{Quantifying the effect of Glasma.} In this study, the main quantity of interest is the momentum broadening, a measure of the momentum accumulated by a particle as it passes through the Glasma. It is computed as $\delta p_i^2(\tau)\equiv p_i^2(\tau)-p_i^2(\tau_\mathrm{form})$ and then averaged over multiple particle trajectories and Glasma background events, yielding $\langle \delta p_i^2(\tau)\rangle$.  Its derivative defines instantaneous transport coefficients at each value of the proper time, namely
\begin{equation}
    \label{eq:kappaqhat}
    \frac{\d }{\d\tau}\langle\delta p^2_i(\tau)\rangle\equiv \begin{cases}\kappa_i(\tau),&\text{heavy quarks,}\quad\text{with }i\in{L, T}\\
    \hat{q}_i(\tau),&\text{jets,}\phantom{\text{vy quarks}}\quad\text{with }i\in{z, y}\end{cases}.
\end{equation}
The heavy quarks are initialized with $p_T(\tau_\mathrm{form})$ and the momentum broadening is computed along the longitudinal $L$ and transverse $T$ directions, evaluated with respect to the beam axis, namely $\langle\delta p_L^2\rangle\equiv \langle\delta p_z^2\rangle$ and $\langle\delta p_T^2\rangle\equiv \langle\delta p_x^2\rangle+\langle\delta p_y^2\rangle$. We choose the jets to propagate along $p_x(\tau_\mathrm{form})$, with longitudinal and transverse momentum broadenings corresponding to $\langle\delta p_z^2\rangle$ and $\langle\delta p_y^2\rangle$. Since the Glasma is anisotropic, we expect the momentum broadening of particles in the Glasma to also be anisotropic. Thus, we extract the ratio of longitudinal over transverse momentum broadenings, namely $\langle\delta p_L^2\rangle\big/\langle\delta p_T^2\rangle$ for heavy quarks and $\langle\delta p_z^2\rangle\big/\langle\delta p_y^2\rangle$ for jets, as a measure of quark momentum anisotropy. 

For particles, we use a toy model initialization. They are randomly created in the transverse plane $\undertilde{x}(\tau_\mathrm{form})$ with $\eta(\tau_\mathrm{form})=0$, at a formation time proportional to the inverse of the mass $\tau_\mathrm{form}\propto 1/(2m)$ for heavy quarks and $\tau_\mathrm{form}=0$ for jets. All particles are formed with a given initial momentum,  $p_T(\tau_\mathrm{form})$ for heavy quarks and $p_x(\tau_\mathrm{form})$ for jets, an input parameter which is then varied. Our Glasma is suitable for central collisions of large nuclei with infinite extent. The saturation momentum is chosen as $Q_s=2\,\mathrm{GeV}$. Both the Glasma and particle have periodic boundary conditions. 

\textbf{Lattice correlators.} In particular quark kinematics limits, the particle and Glasma equations decouple and the momenta evolution from Eq.~\eqref{eq:wong} enables one to extract the momentum broadening solely from Glasma lattice field correlators. These cases correspond to infinitely massive and highly energetic quarks. For the very massive static quarks ${\color{customblue}m\rightarrow\infty}$, the momentum broadening is extracted from Glasma electric field correlators as \cite{Avramescu:2023qvv}
\begin{equation}
    \label{eq:static}
    \big\langle\delta p_i^2(\tau)\big\rangle_{{\color{customblue}m\rightarrow\infty}}^\mathrm{static}=g^2 \int\limits_0^{\tau}\mathrm{d}\tau^\prime\int\limits_0^{\tau}\mathrm{d}\tau^{\prime\prime}\,\big\langle\tr{{\color{customblue}E_i}(\tau^\prime){\color{customblue}E_i}(\tau^{\prime\prime})}\big\rangle.
\end{equation}
In the case of very fast lightlike jets propagating along the $x$-direction with ${\color{customred}p_x\rightarrow\infty}$, the momentum broadening is extracted from parallel transported electromagnetic fields according to \cite{Ipp:2020nfu}
\begin{equation}
    \label{eq:lightlike}
    \big\langle \delta p_i^2(\tau)\big\rangle_{{\color{customred}p_x\rightarrow\infty}}^\mathrm{lightlike}=g^2 \int\limits_0^\tau \d \tau^{\prime} \int\limits_0^\tau \d \tau^{\prime \prime}\big\langle\mathrm{Tr}\{{\color{customred}\widetilde{F}_i}(\tau^{\prime}) {\color{customred}\widetilde{F}_i}(\tau^{\prime \prime})\}\big\rangle.
\end{equation}
The color force components ${\color{customred}F_x}\equiv E_x, \, {\color{customred}F_y}\equiv E_y-B_z, \, {\color{customred}F_z}\equiv E_z+B_y$ are parallel transported ${\color{customred}\widetilde{F}_i}\equiv \mathcal{U}_x^\dagger{\color{customred}F_i}\,\mathcal{U}_x$ with Glasma Wilson lines along the $x$-direction. The momentum broadening and transport coefficients derived from lattice correlators provide valuable numerical checks and a baseline of comparison for dynamical quarks simulated with our solver. 

\begin{figure}[!hbt]
\includegraphics[width=\textwidth]{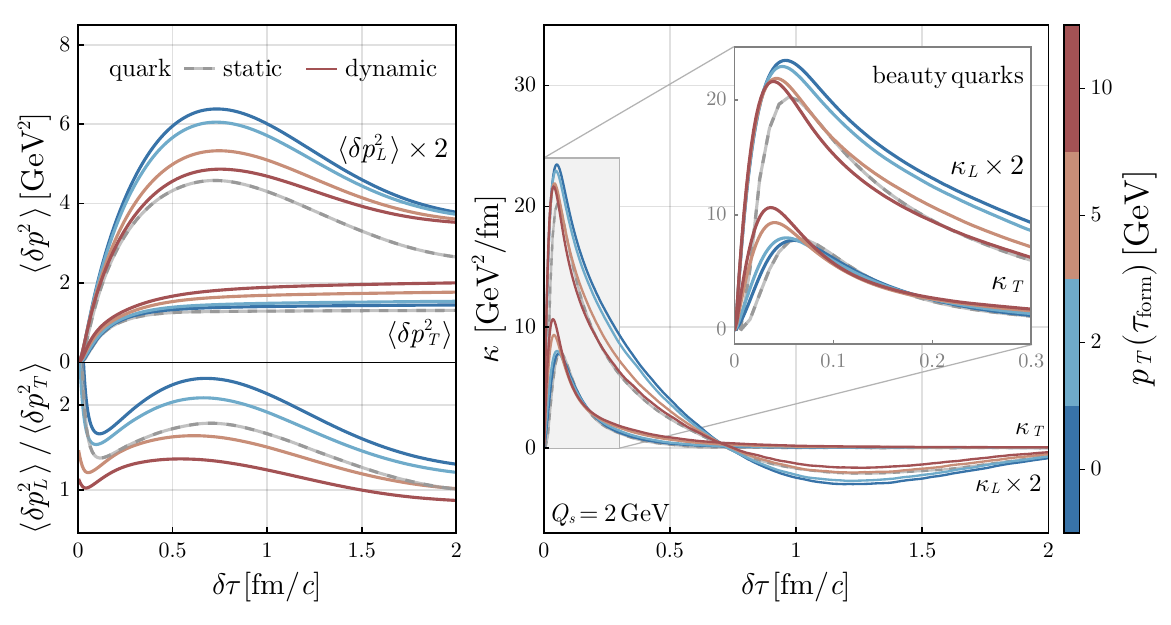}
\caption{\label{fig:wong_kappa} Proper time evolution in terms of $\delta\tau\equiv \tau-\tau_\mathrm{form}$ of \textit{(top left)} longitudinal \textit{(multiplied by 2)} and transverse components of the momentum broadening, \textit{(bottom left)} their ratio which yields the anisotropy \textit{(right)} instantaneous transport coefficients $\kappa$, both longitudinal \textit{(multiplied by 2)} and transverse with \textit{(right inset)} a zoom-in on the very-early stage. All simulations are done for beauty quarks with $m_\mathrm{beauty}\approx 4.2\,\mathrm{GeV}$. We compare dynamical quarks, as simulated using Wong's equations and having fixed  $p_T(\tau_\mathrm{form})$ \textit{(colored lines)} with static quarks \textit{(grey dashed lines)} extracted from lattice Glasma electric fields correlators as in Eq.~\eqref{eq:static}.}
\end{figure}

\textbf{Heavy quarks.} We extract both the longitudinal $\langle \delta p^2_L\rangle$ and transverse $\langle \delta p^2_T\rangle$ momentum broadenings of dynamical beauty quarks with initial $p_T(\tau_\mathrm{form})\in\{0,2,5,10\}\,\mathrm{GeV}$ and compare them to the static quark case from Eq.~\eqref{eq:static}. The dynamical quarks exhibit deviations from the infinite quark mass scenario, as shown in all the curves from Fig.~\ref{fig:wong_kappa}. The quark momentum is anisotropic throughout the evolution, see \textit{lower left} of Fig.~\ref{fig:wong_kappa} and the anisotropy is ordered with the inverse of the initial $p_T$. Interestingly, static quarks which have null momentum are less anisotropic than dynamical quarks with vanishing initial $p_T$ but which propagate in the Glasma fields. On the other hand, faster quarks with larger initial $p_T$ become less anisotropic than the static ones.

Both $\langle \delta p^2_L\rangle$ and $\langle \delta p^2_T\rangle$ increase rapidly at extremely early times, see \textit{top left} of Fig.~\ref{fig:wong_kappa}, when the heavy quarks are rapidly accelerated by the coherent electric Glasma flux tubes. As the Glasma fields expand longitudinally and become dilute, this increase is slowed down. These cause the corresponding derivatives $\kappa_L$ and $\kappa_T$ to exhibit a very large initial peak, followed by a gradual decrease, as represented in the \textit{right inset} of Fig.~\ref{fig:wong_kappa}. The total $\kappa_\mathrm{peak}\approx 19-22\,\mathrm{GeV^2/fm}$ depends on the initial $p_T$ and is located at $\delta\tau_\mathrm{peak}\approx 0.03-0.05\,\mathrm{fm/}c$. At late times this increase either plateaus for $\langle \delta p^2_T\rangle$, yielding a null $\kappa_T$ or intriguingly decreases for $\langle \delta p^2_L\rangle$, giving a negative $\kappa_L$. A possible explanation would be that the longitudinal dynamics are triggered by plasmon oscillations of the Glasma fields, as similarly reported in \cite{Boguslavski:2020tqz}. Moreover, the heavy quark $\kappa$ in the Glasma is fundamentally different from the diffusion coefficient obtained in a Langevin approach. There, $\kappa$ increases linearly without reaching a peak, is less steep and has smaller values than in Glasma.

\begin{figure}[!hbt]
\includegraphics[width=\textwidth]{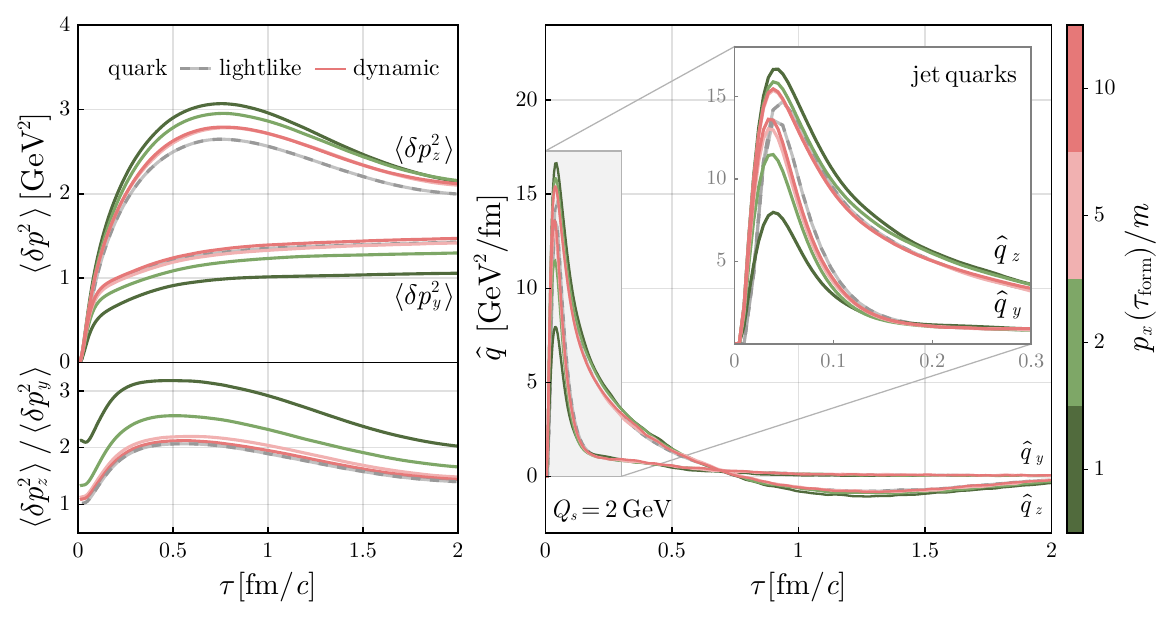}
\caption{\label{fig:wong_qhat} \textit{(Top left)} Momentum broadening along longitudinal $z$ and in transverse $y$ directions, \textit{(bottom left)} jet momentum anisotropy given by their ratio, \textit{(right)} jet directional transport coefficients $\hat{q}_z$ and $\hat{q}_y$ with \textit{(right inset)} a zoom-in on extremely early times, as a function of proper time. The simulations are done for fast quarks, formed at $\tau_\mathrm{form}=0\,\mathrm{fm/}c$, with various initial momenta $p_x(\tau_\mathrm{form})$ and masses $m$, while the results are plotted in terms of their ratio $p_x(\tau_\mathrm{form})/m$ \textit{(colored lines)}. These are compared to the lightlike quarks \textit{(grey dashed lines)} computed from lattice Glasma electric and magnetic field correlators, see Eq.~\eqref{eq:lightlike}.}
\end{figure}

\textbf{Jets.} We simulate dynamical jet quarks with fixed initial $p_x(\tau_\mathrm{form})\in\{1, 2, 5, 10\}$, extract the longitudinal $\langle \delta p^2_z\rangle$ and transverse $\langle \delta p^2_y\rangle$ momentum broadenings, and then compare them to the lightlike jet results evaluated with Eq.~\eqref{eq:lightlike}. In a similar manner as for heavy quarks, dynamical jet curves differ from the limiting case of extremely fast quarks, as may be seen in Fig.~\ref{fig:wong_qhat}. The anisotropy for dynamical jets is consistently larger than the lightlike jets, see \textit{(bottom left)} part of Fig.~\eqref{fig:wong_qhat}, but both results are highly anisotropic at the typical times when the Glasma stage would end, roughly $\delta\tau_\mathrm{Glasma}\approx 0.3\,\mathrm{fm/}c$. The proper time evolution of the longitudinal and transverse jet momentum broadening is parametrically compatible with the heavy quarks, see Fig.~\ref{fig:wong_kappa}, since similar underlying Glasma dynamics govern the very early and late time behaviors. Consequently, the jet transport coefficients $\hat{q}_z$ and $\hat{q}_y$ have a sharp increase, with a peak of total $\hat{q}_\mathrm{peak}\approx 24-30\,\mathrm{GeV^2/fm}$, depending on mass and initial momentum but located at the same proper time $\tau_\mathrm{peak}\approx 0.04\,\mathrm{fm/}c$, followed by a long-lasting decay. Similarly, at late times, the transverse $\hat{q}_y$ is null while $\hat{q}_z$ becomes negative, being subject to the same first Glasma plasmon oscillation. Curiously, as we report in \cite{Avramescu:2023qvv}, these longitudinal momentum broadening oscillations continue to persist only for heavy quarks. 

\section*{Summary}  We developed a numerical solver for heavy quarks and jets in the Glasma fields. We extracted the momentum broadening and used it to compute transport coefficients, namely $\kappa$ and $\hat{q}$. In the Glasma, they are large, steep and exhibit a peak. Moreover, the accumulation of momentum is anisotropic. All of these quantities vary with respect to the details of particle initialization, namely mass, formation time and initial momentum. 

\textbf{Acknowledgments.} D.~A.~acknowledges funding from the Academy of Finland, Center of Excellence in Quark Matter project 346324. V.~G.~acknowledges funding from UniCT under ``Linea di intervento 2'' (HQCDyn Grant). D.~M.~acknowledges funding from the Austrian Science Fund (FWF) projects P~34455 and P~34764. We are grateful to K. Boguslavski, T. Lappi and H. Mäntysaari for many insightful discussions.

\bibliographystyle{JHEP}
\bibliography{references}

\end{document}